\documentclass[aps,pra,preprint,groupedaddress]{revtex4-1}

\usepackage{graphicx}
\begin{document}


\title{Second Harmonic Enhanced Feedforward Intensity Noise Eater}


\author{M. Stefszky and C. Silberhorn}
\affiliation{Integrated Quantum Optics, Applied Physics, University of Paderborn, Warburger Strasse 100, 33098 Paderborn, Germany}


\date{\today}

\begin{abstract}
Quantum nondemolition theory has been well understood for a number of decades, however, applications of such techniques remain limited owing to the increased complexity that these techniques require. In this paper, quantum nondemolition theory is used to investigate the performance of a real world device, an electro-optic feed-forward intensity noise eater. It is shown that by replacing the typical beamsplitter in such a device with a single pass second harmonic generation followed by a dichroic mirror the performance of the noise eater can be significantly improved, even with low conversion efficiencies. 
\end{abstract}

\pacs{}

\maketitle


\section{Introduction}

Quantum nondemolition (QND) measurements were first introduced as a method of extracting the most amount of information from a given system, without adding additional noise to following measurements \cite{Braginsky80.SCI, Grangier98.N}. An ideal QND measurement can only be made when the observable in question $\hat{A}$, commutes with itself at different moments in time, $[\hat{A}(t_0),\hat{A}(t_1)]=0$. QND measurement theory, however, provides a way of quantifying nonideal systems by evaluating the disturbance made by a measurement and the amount of information gained through the measurement.

Although QND theory is well established, there are still few situations in which QND devices are routinely used. This is in part due to the fact that QND measurements typically require complex setups in order to provide the desired benefits. Shapiro {\it et al} \cite{Shapiro80.OL} showed that squeezing of the meter field in a simple beamsplitter setup is enough to provide QND enhancement. This technique is one of the simpler methods for achieving QND measurements, but nevertheless requires the (nontrivial) production of an optical squeezer in order to produce the required squeezed meter field.

Another method that has been investigated is second harmonic generation (SHG) \cite{Olsen00.PRA, Dance91.PRL}. Second harmonic generation has the advantage of generally being easier to experimentally implement than a squeezer. The main problem with SHG QND is that reaching the QND regime requires very large conversion efficiencies that are, at the very least, difficult to achieve in experimental setups. Additionally, the high conversion second harmonic process removes the majority of the power from the original field, an undesirable effect in many applications. 

However, although second harmonic generation does not fulfil all the QND criteria at low conversion efficiencies, it is known that the properties of the output fields are strongly altered by the nonlinear interaction \cite{Ou94.PRA}. Weak levels of optical squeezing occur, and the correlations between the various fields can be much stronger than those seen in the case of a beamsplitter. Furthermore, feedforward and feedback devices in quantum nondemolition theory have been investigated previously \cite{Buchler99.OL,Shapiro87.JOSAB,Wiseman94.PRA}. 

In this paper we investigate a new method for achieving QND enhancement of a feedforward scheme. By simply replacing the beamsplitter component of a standard feedforward noise eater scheme with a second harmonic generation stage one can show a substantial improvement in device performance, {\it even in low conversion regimes where the typical QND measures are not met}.

\section{The Feedforward Intensity Noise Eater}

We begin by clearly defining the goal of the intensity noise eater. The aim of an intensity noise eater is typically considered to be reduction of the the amplitude quadrature noise on an initial noisy laser beam, which we label the input signal field. However, one also has to consider the power in the initial beam as a resource for most applications. Therefore the aim of the device that will be used throughout this paper is to reduce the noise on the input field to some desired level, {\it whilst retaining as much of the initial power as possible}.

\begin{figure}[!h]
  \caption{(Color online) The standard beamsplitter (BS) noise eater. A bright, noisy, initial field is incident on a beamsplitter. The reflected portion of the light, the meter field, is detected and the information on this field is used to reduce the noise on the remaining field through the use of an electro-optic intensity modulator.}
  \centering
    \includegraphics[width=0.6\textwidth]{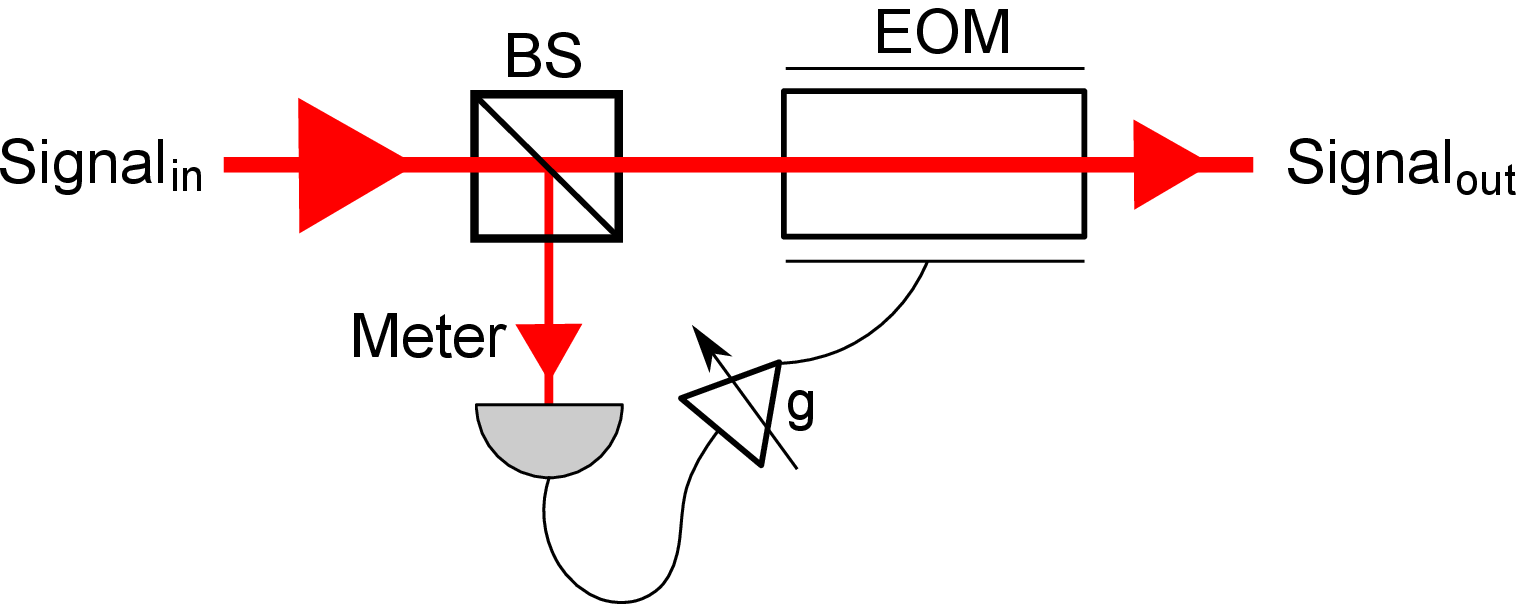}
    \label{NoiseEater}
\end{figure}

In the standard feedforward noise eater setup, which we shall call the beam splitter (BS) noise eater, some percentage of the laser field power is split from the input (signal) field using a beamsplitter and is then detected. We refer to this process as the tap-off and the tap-off field as the meter field. The information on the detected field is then used, via feedforward through an electro-optic modulator, to reduce the noise in the transmitted field. The basic setup is shown in Figure \ref{NoiseEater}


In order to improve upon the BS noise eater design, one might consider a device that shows stronger correlations between the meter and output fields, without introducing additional noise. Perhaps the first most obvious process for examination is second harmonic generation. It is known that SHG can produce squeezed states of light, thereby reducing the output noise before even applying any feedforward, and the correlations between the various fields have been shown to be much stronger than those present after beamsplitting \cite{Ou94.PRA}. In addition, it is experimentally simple to replace the beamsplitter in a noise eater setup with second harmonic generation in single-pass followed by a dichroic mirror for separation of the output signal and meter fields. Furthermore, the increase in loss involved in such a transition can be very minor.

\begin{figure}[!h]
  \caption{(Color online) The second harmonic (SH) noise eater. A bright, noisy, initial field is incident on a second harmonic generation device, here assumed to be single-pass. A dichroic mirror is placed after the SH stage such that the SH field, acting as the meter field, and the input signal field are separated. As before, the information on the meter field is used to reduce the noise on the output signal field.}
  \centering
    \includegraphics[width=0.6\textwidth]{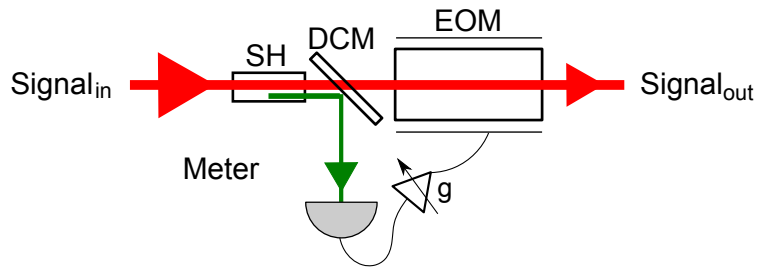}
    \label{SHNoiseEater}
\end{figure}

The new setup, which we shall call the second harmonic intensity noise eater (SH noise eater), is shown in figure \ref{SHNoiseEater}. In the following sections we compare this system, the SH noise eater, and the standard noise eater setup.

\section{Theory}

The system is investigated using standard QND theory techniques \cite{Holland90.PRA,Grangier98.N,Smith93.OC}. We work with linearised operators in the frequency domain and therefore the validity of the results presented here reduces towards very high conversion efficiencies (above 99\%) \citep{Olsen00.PRA}. However, this is not a concern because we are primarily interested in the performance at low conversion efficiencies.

\begin{figure}[!h]
  \caption{(Color online) The relevant fields in the general feedforward intensity noise eater scheme. Variables described in text.}
  \centering
    \includegraphics[width=0.6\textwidth]{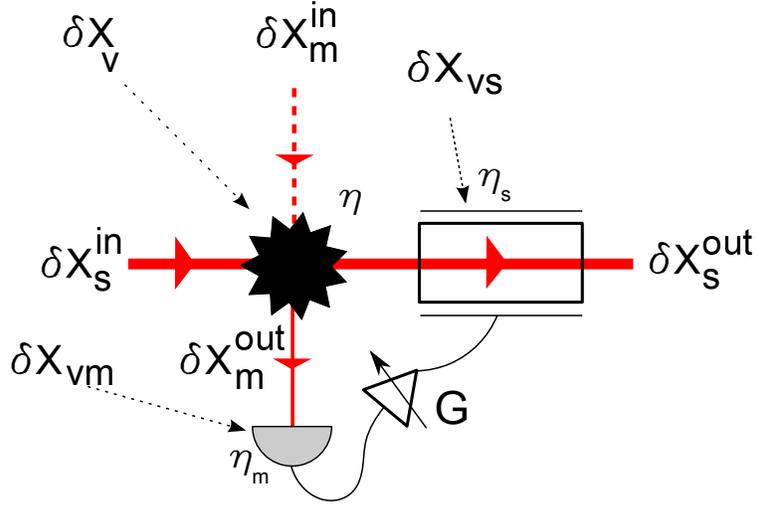}
    \label{FBB}
\end{figure}

The setup that describes the transfer of amplitude quadrature noise fluctuations $\delta{\rm  \tilde{X}}$, is shown in Figure \ref{FBB}. A bright input signal field with fluctuations ${\rm \delta \tilde{X}_{s}^{in}}$, and a vacuum input meter field ${\rm \delta \tilde{X}_{m}^{\rm in}}=1$, undergo some tap-off process that maps the power of the input signal field and the fluctuations from both input fields to the two outputs. The noise on the output meter and signal fields are given ${\rm \delta \tilde{X}_{m}^{out}}$, and ${\rm \delta \tilde{X}_{s}^{out}}$, respectively. The meter field is then measured on a photodetector with efficiency ${\rm \eta_{m}}$. This measured signal is then amplified with a variable gain of G and fed forward to the output signal field via an electro-optic intensity modulator. Vacuum terms can enter via losses in the tap-off process (due to imperfect coatings or absorption) ${\rm \delta \tilde{X}_{v}}=1$, the meter arm due to imperfect detection ${\rm \delta \tilde{X}_{vm}}=1$, and the signal arm due to losses introduced by the intensity modulator, ${\rm \delta \tilde{X}_{vs}}=1$.

The final parameter is the tap-off ratio $\eta$, given by the ratio of the power in the output signal field to the power in the input signal field. For a beamplitter, this value is equal to the beamsplitter transmission, but for the SHG process it is related to the nonlinear conversion efficiency of the SHG device. In the case of SHG, this is experimentally equivalent to changing the length of the nonlinear medium but is also a function of the power in the signal field.

We first define the transfer matrix of the tap-off process. We can write

\begin{eqnarray}
\left( \begin{array}{c}
\delta {\rm \tilde{X}_{s}^{to}} \\
\delta {\rm \tilde{X}_{m}^{to}}
\end{array} \right)
& = &
\left( \begin{array}{c c c}
a	&	b	&	c \\
d	&	e	&	f
\end{array} \right)
\left( \begin{array}{c}
\delta {\rm \tilde{X}_{s}^{in}} \\
\delta {\rm \tilde{X}_{m}^{in}} \\
\delta {\rm \tilde{X}_{vl}^{in}}
\end{array} \right),
\end{eqnarray}
where the terms $\delta {\rm \tilde{X}_{s}^{to}}$ and $\delta {\rm \tilde{X}_{s}^{to}}$ describe the amplitude quadrature fluctuations of the signal and meter fields immediately after the tap-off process. 

For the lossless beamsplitter the process transfer matrix is well known and is given by

\begin{eqnarray}
\left( \begin{array}{c c c}
a	&	b	&	c \\
d	&	e	&	f
\end{array} \right)_{BS} & = &
\left( \begin{array}{c c c}
\sqrt{\eta}		&	-\sqrt{1-\eta}	&	0 \\
\sqrt{1-\eta}	&	\sqrt{\eta}		&	0
\end{array} \right),
\end{eqnarray}
Note here that we have assumed that the beamsplitter is lossless. This assumption is valid because the losses in the beamsplitter are small. 

Likewise, the terms for the second harmonic process can be found from the lossless squeezer Hamiltonian \cite{Li94.PRA,Ou94.PRA}. Here we model a single-pass experiment, allowing us to assume a lossless SH process. Were a resonant SH process to instead be used, then the noise coupling terms, $c$ and $f$, would have to be included \cite{Bruckmeier97.PRL}. The second harmonic transfer matrix is then given by

\begin{eqnarray}
\left( \begin{array}{c c c}
a	&	b	&	c \\
d	&	e	&	f
\end{array} \right)_{SH} & = &
\left( \begin{array}{c c c}
\left(1-\xi\,{\rm tanh}\,\xi \right){\rm sech}\,\xi		& \sqrt{2}\,{\rm tanh}\,\xi\,{\rm sech}\,\xi	&	0 \\
-\left({\rm tanh}\,\xi+\xi\,{\rm sech}^{2}\,\xi \right)/\sqrt{2}	&	{\rm sech}^{2}\,\xi				&	0
\end{array} \right),
\end{eqnarray}
where $\xi$ is the normalized interaction strength, which is proportional to the nonlinear interaction strength, the fundamental field magnitude and the length of the nonlinear medium.


Now that the transfer matrices for the idealised tap-off process are known, the effect of the feedforward and losses can be added to complete the theoretical description. We follow the same method described by Buchler {\it et al.} \cite{Buchler99.PRA} to arrive at

\begin{eqnarray}
\left( \begin{array}{c}
\delta {\rm \tilde{X}_{s}^{out}} \\
\delta {\rm \tilde{X}_{m}^{out}}
\end{array} \right)
& = &
\left( \begin{array}{c c}
\sqrt{\eta_{\rm s}}	&	G \sqrt{\eta_{\rm s} \eta_{\rm m}} \\
0				&	\sqrt{\eta_{\rm m}}
\end{array} \right)
\left( \begin{array}{c c c}
a	&	b	&	c \\
d	&	e	&	f
\end{array} \right)
\left( \begin{array}{c}
\delta {\rm \tilde{X}_{s}^{in}} \\
\delta {\rm \tilde{X}_{m}^{in}} \\
\delta {\rm \tilde{X}_{vl}^{in}}
\end{array} \right) \nonumber \\ 
& &
+\left( \begin{array}{c c}
\sqrt{1-\eta_{\rm s}}	&	{\rm G} \sqrt{\eta_{\rm s} \left( 1-\eta_{\rm m} \right)} \\
0					&	\sqrt{1-\eta_{\rm m}}
\end{array} \right)
\left( \begin{array}{c}
{\rm \delta \tilde{X}_{vs}} \\
{\rm \delta \tilde{X}_{vm}}
\end{array} \right),
\end{eqnarray}
where G is the feedforward gain, which can be complex. The two loss terms, $\eta_s$ and $\eta_m$, represent the losses in the signal output and meter fields respectively. For the meter arm $\eta_m$ is equivalent to the quantum efficiency of the photodetector and for the signal arm $\eta_s$ is the transmission of the loss equivalent beamsplitter that includes losses due to the modulator and the tap-off process.

\section{Tap-Off Process}

Before looking at the results for the full feedforward configuration, it is informative to first directly compare the performance of the two idealised tap-off processes. We set the feedforward gain to zero ($G=0$) and ignore losses from the modulator and meter detection ($\eta_m=\eta_s = 1$). We then compare the BS and SH processes as the the tap-off ratio $\eta$, is varied.

Different QND measures can be used to characterise the fields after the tap-off process. The measures we will use here are the output variance, the information transfer coefficients, the conditional variance and the correlations. Each of these are explored in the following subsections before reintroducing the feedforward and directly comparing the performance of the two system as a noise eater.

\subsection{Output Variances}

\begin{figure}[!h]
  \caption{(Color online) Variance of output signal fields as the amount of power in the tap-off is increased for the SH and BS processes on noisy input fields. The grey region indicates squeezing. Parameters are ${\rm V_{s}^{in}}=10$, G=0, ${\rm \eta_{m}} = 1$, ${\rm \eta_{s}} = 1$.}
  \centering
    \includegraphics[width=0.6\textwidth]{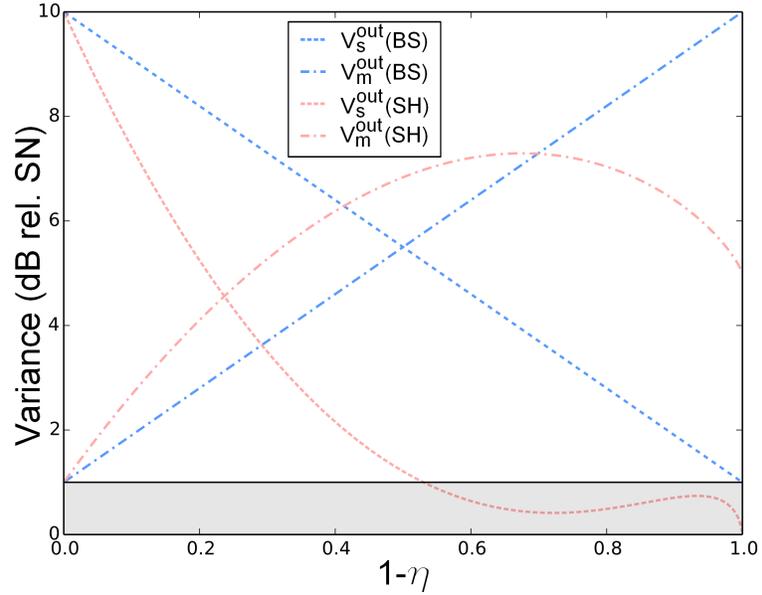}
    \label{TferV}
\end{figure}

The first measure that we investigate is the variance $V = \langle\delta {\rm \tilde{X}}^2\rangle$, of the two output fields. It has been shown many times that the output of a second harmonic process is capable of squeezing the output modes \cite{Li94.PRA,Ou94.PRA,Olsen00.PRA}. We assume that the input state has noise that is ten times above the shot noise ($\delta {\rm \tilde{X}_{s}^{in}}=\sqrt{10}$). Figure \ref{TferV} shows the output variances of the fields as the tap-off ratio is varied. We see that in contrast to the BS setup, the SH setup has more information on the meter field, and less noise on the output signal field. This is exactly the desired behaviour provided that the correlations between the meter and the signal are high (which will be shown in section \ref{corsec}).

\subsection{Information Transfer Coefficients}
The next measure that we define is the information transfer coefficient. We define the meter field signal transfer coefficient ${\rm T_{m}}$, the signal field information transfer coefficient${\rm T_{s}}$, and a total information transfer coefficient ${\rm T_{s+m}}$, for the BS setup

\begin{eqnarray}
{\rm T_{s} = \frac{SNR_{s}^{out}}{SNR_{s}^{in}}}, \\
{\rm T_{m} = \frac{SNR_{m}^{out}}{SNR_{m}^{in}}}, \\
{\rm T_{s+m} = T_{s}+T_{m} },
\end{eqnarray}

where the signal to noise ratio (SNR) of the fields is the ratio between the measured signal strength, ${\rm (S)}$, with the noise component subtracted, to the noise level (in this case shot noise) ${\rm N}$,

\begin{eqnarray}
{\rm SNR = \frac{S-N}{N}},
\end{eqnarray}
where N is the noise due to the quantum limit and S is the measured variance. For values of $1<{\rm T_{s+m}}<2$ one the system is said to be operating as a quantum-optical tap \cite{Goobar93.PRL,Shapiro80.OL}.


Once again it appears as thought he SH setup is advantageous. The transfer coefficient for the meter field is always higher in the SH setup than in the BS and the transfer coefficient on the output signal field is lower than for the BS until very high tap-off ratios are reached.


\subsection{Conditional Variance}

The conditional variance is another measure used to classify the performance of a QND measurement. The conditional variance, ${\rm V_{s|m}}$, is a measure of how much information one gains about the output signal field by measuring the meter field. A conditional variance of less than 1 ,${\rm V_{s|m}} \geq 1 $, indicates non-classical behaviour and the process is regarded as achieving quantum state preparation\cite{Bruckmeier97.PRL,Goobar93.PRL}. The conditional variance between the output signal field and the meter field is given by

\begin{eqnarray}
{\rm V_{s|m} = V_{s}^{out}-\frac{|\left\langle \delta \hat{X}_{s}^{out} \hat{X}_{m}^{out} \right\rangle|^{2}}{V_{m}^{out}}}.
\end{eqnarray}

\begin{figure}[!h]
  \caption{(Color online) Conditional variance for BS and SH processes as the amount of power in the tap-off is increased. The grey region indicates quantum state preparation. Parameters are ${\rm V_{s}^{in}}=10$, G=0, ${\rm \eta_{m}} = 1$, ${\rm \eta_{s}} =1$.}
  \centering
    \includegraphics[width=0.6\textwidth]{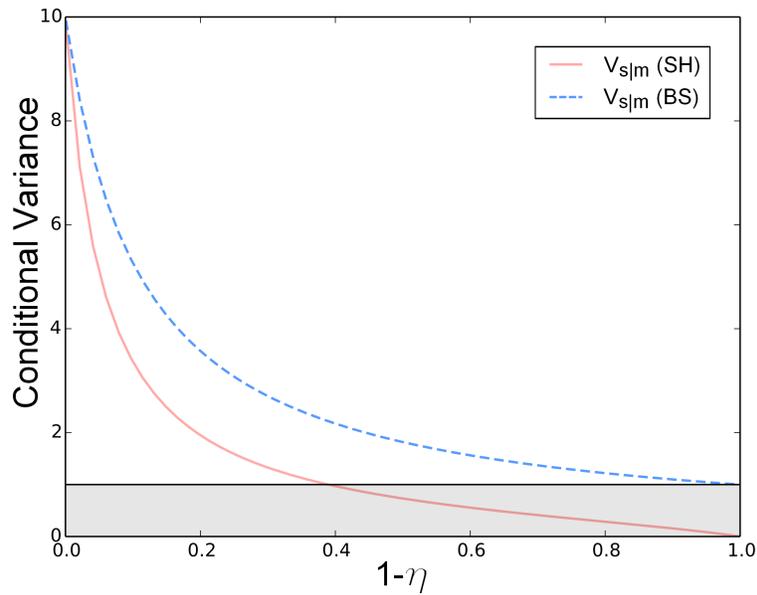}
    \label{CondV}
\end{figure}

Figure \ref{CondV} shows the conditional variance between the output signal field and the meter field. We see that the conditional variance of the SHG process is less than that for the beamsplitter at all splitting ratios $\eta$, as desired. We also note that for some tap-off ratios, the process fulfills quantum state preparation requirements.

\subsection{Correlations}  \label{corsec}

The final measure is the correlations between the fields. The noise eater requires strong correlations between the meter field and the output signal field to operate effectively. The correlations between two fields, 1 and 2, can be written

\begin{eqnarray}
{\rm C_{1,2} = \frac{|\left\langle \delta \hat{X}_{1} \delta \hat{X}_{2} \right\rangle|^{2}}{V_{1}{V_{2}}}}. \label{CorEq}
\end{eqnarray}

\begin{figure}[!h]
  \caption{(Color online) Correlation functions for BS and SH tap-off processes as the amount of power in the meter field is increased. Parameters are ${\rm V_{s}^{in}}=10$, G=0, ${\rm \eta_{m}} = 1$, ${\rm \eta_{s}} = 1$.}
  \centering
    \includegraphics[width=0.6\textwidth]{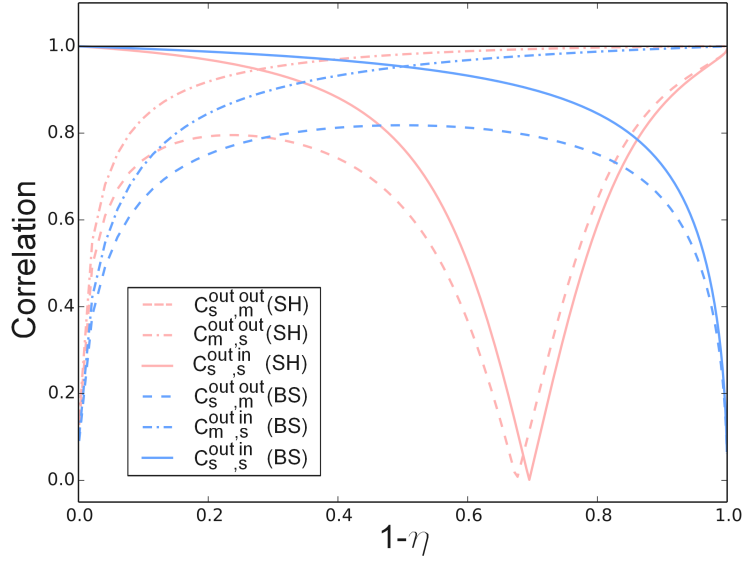}
    \label{CorF}
\end{figure}

The correlations between the fields are illustrated in Figure \ref{CorF}. We see that the correlations between the output signal and the output meter fields for the SH process are greater than for the beamsplitter up until some point where the two values cross (for this case at around $1-\eta = 0.3$). This indicates that feedforward will provide the largest reduction in the output noise at lower tap-off percentages.

\section{Noise Eater}

\begin{figure}[!h]
  \caption{(Color online) Variances of relevant output fields in dB relative to shot noise for the SH and BS noise eaters, as the feedforward gain is varied. Black (solid) lines are added to highlight the minimum noise levels obtained for the two cases. Parameters are $\eta = 0.9 $, ${\rm V_{s}^{in}}=10$, ${\rm \eta_{m}} = 0.9$, ${\rm \eta_s^{BS}} = 0.95$, ${\rm \eta_s^{SH}} = 0.95$.}
  \centering
    \includegraphics[width=0.6\textwidth]{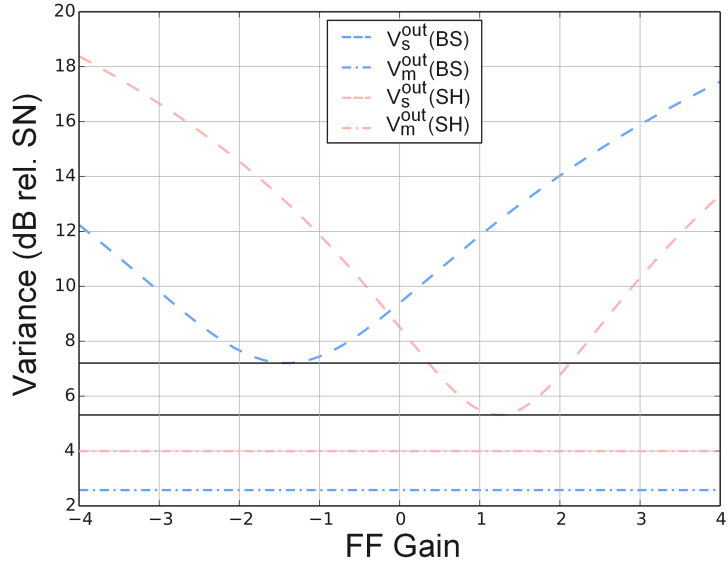}
    \label{OutV}
\end{figure}

Now that the two tap-off processes are well understood, we turn our attention to the full noise eater system. To model the noise eater we simply switch on the feedforward gain, G. We choose a tap-off value of 10 percent ($\eta=0.9$) and look at the output variance of the fields as the gain (assumed to be real) is varied in both the SH and BS setups.  In order to make the comparison clearer, we assume that both the second harmonic process and the beamsplitter introduce the same loss of 5\% ($\eta_s^{BS} = \eta_s^{SH} = 0.95$) and that the modulator also has a loss of 5\% ($\eta_s = 0.9$). The detector in the meter arm has an assumed quantum efficiency of 90\% ($\eta_m = 0.9$). The variance of the output signal field and the measured meter field are plotted for both the SH and BS noise eater systems in Figure \ref{OutV}. We immediately see that the minimum noise of the SH noise eater is approximately 2$\,$dB below the minimum noise level reached by the BS noise eater, and therefore the SH noise eater is clearly outperforming the BS noise eater.

\begin{figure}[!h]
  \caption{(Color online) The variances of the output signal fields for SH noise eater and BS noise eater for optimum gain with varying tap-off ratio. The light grey shaded region indicates operating parameters for which the device fulfils the "quantum state preparation" criterion, and the dark shaded region indicates the region where the device fulfils the "QND measurement" criterion. The black trace illustrates the difference, in dB, between the BS and SH noise eaters. Parameters are ${\rm V_{s}^{in}}=10$, ${\rm \eta_{m}} = 0.9$, ${\rm \eta_s^{BS}} = 0.95$, ${\rm \eta_s^{SH}} = 0.95$.}
  \centering
    \includegraphics[width=0.6\textwidth]{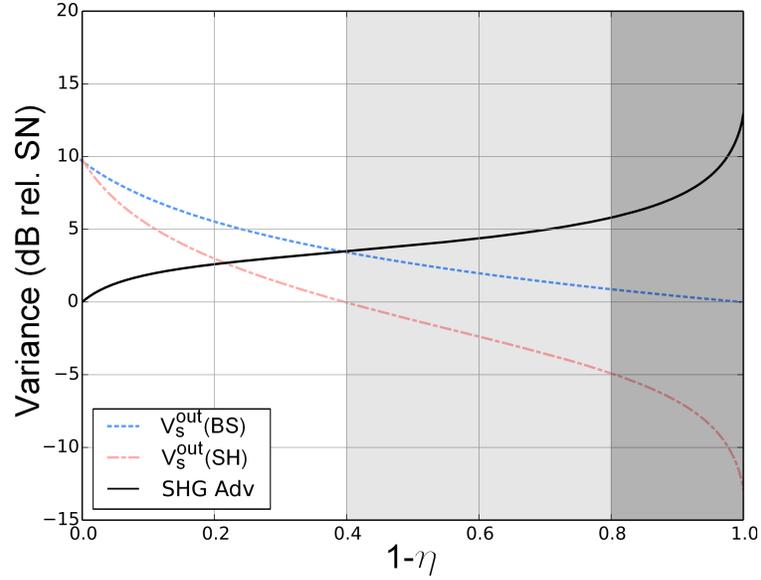}
    \label{VaryG}
\end{figure}

Finally, we investigate how the SHG noise eater compares for all values of the splitting between the meter and signal output fields. In Figure \ref{VaryG}, a search over the (real) feedforward gain is performed for each value of the tap-off ratio. The minimum output variance of the output signal field found through each of these searches is plotted. It is immediately apparent that the SH noise eater clearly outperforms the BS noise eater, \textit{even in regions where the device does not fulfil any of the standard quantum nondemolition criteria.}

\section{Conclusion}

We have shown that a noise eater setup in which the standard beamsplitter is replaced with a second harmonic process is capable of substantially outperforming the original device. This is true even at low conversion efficiencies, where the SHG process does not fulfil any of the standard QND measurement criteria. Although the benefits gained will depend on many factors such as operational powers, losses, and specific laser systems used, the results show that consideration of the SH enhanced noise eater may lead to a device with much improved performance for very little increase in complexity.

\begin{acknowledgments}
The authors acknowledge funding from the Gottfried Wilhelm Leibniz-Preis.
\end{acknowledgments}

\bibliography{2014JulBib}

\clearpage

\begin{figure}[!h]
  \caption{(Color online) Information transfer for BS and SH tap-off processes as the amount of power in the tap-off is increased. The grey region indicates parameters where the total information transfer coefficient indicates non-classical (quantum-optical tap) behaviour. Parameters are ${\rm V_{s}^{in}}=10$, G=0, ${\rm \eta_{m}} = 1$, ${\rm \eta_{s}} = 1$.}
  \centering
    \includegraphics[width=0.6\textwidth]{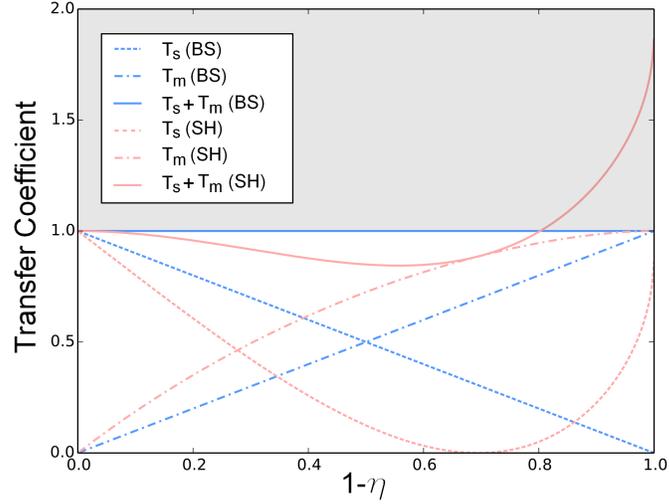}
    \label{ITC}
\end{figure}

%

\end{document}